# Topology, homogeneity and scale factors for object detection: application of eCognition software for urban mapping using multispectral satellite image

**Lemenkova Polina**
**Faculty of Science, Institute for Environmental Studies**
**Charles University in Prague,**
**Prague, Czech Republic**
pauline.lemenkova@gmail.com

*Abstract*—The research scope of this paper is to apply spatial object based image analysis (OBIA) method for processing panchromatic multispectral image covering study area of Brussels for urban mapping. The aim is to map different land cover types and more specifically, built-up areas from the very high resolution (VHR) satellite image using OBIA approach. A case study covers urban landscapes in the eastern areas of the city of Brussels, Belgium. Technically, this research was performed in eCognition raster processing software demonstrating excellent results of image segmentation and classification. The tools embedded in eCognition enabled to perform image segmentation and objects classification processes in a semi-automated regime, which is useful for the city planning, spatial analysis and urban growth analysis. The combination of the OBIA method together with technical tools of the eCognition demonstrated applicability of this method for urban mapping in densely populated areas, e.g. in megapolis and capital cities. The methodology included multiresolution segmentation and classification of the created objects.

*Keywords*-Urban mapping, object based image analysis

## I. Introduction

The satellite imagery is the most applicable source of information for urban mapping. It enables detecting single buildings within the whole land cover pattern.

Combination of various types of classification of the remote sensing data focused on the spectral, spatial and textural features of the objects increases the variety of urban mapping methods as well as their accuracy and precision. Some advances have been reported on urban sprawl mapping and assessment [3], [6], [17], [18], [20], [26]. Mapping land cover types using very high resolution (VHR) satellite images is efficient using object based image analysis (OBIA) approach. However, better possibilities and perspectives comparing to traditional cartographic ways of mapping can be achieved from object oriented image approach using smart mapping and *a priori* knowledge.

The core idea of the *a priori* knowledge approach consists in the application of the additional knowledge that is to be applied for the advanced image analysis. Such information can exploit the richness of the object features that is embedded in their properties. For instance, while interpreting field areas on the multispectral satellite image, we may use additional knowledge on urban landscape distribution over the target area in order to detect the nature of the land cover type. In other words, using knowledge based approach we can concentrate on the image segments using our understanding and recognition of the objects instead of the pixels.

The application of the knowledge to the raw data is done by the inference mechanism or engine using "if-then" – a logical, rule based knowledge system [21]. The most easy is to derive geometric information while topological information is sometimes not available. Various works have been published so far explaining OBIA in details [5], [7], [24]. Major approaches used and reported recently [15]. The methodological approaches vary significantly and include such techniques as pixel-based (image classification, regression, etc.), sub-pixel based (linear spectral mixing, imperviousness as the complement of vegetation fraction etc.), object-oriented algorithms, and artificial neural networks [25]. However, though remote sensing data have been used in urban studies many times before, these are often limited by the choices made in data selection: scales, resolutions, quality and spatial extent, as well as vary much in methodology techniques [4], [8], [19], [22], [23], [27]. As urban sprawl and in city land cover types change within the city face, some research were successfully

VII საერთაშორისო სამეცნიერო-პრაქტიკული კონფერენცია ინსო-2015
VII INTERNATIONAL SCIENTIFIC AND PRACTICAL CONFERENCE INSO-2015

reported analysis of the city development using remote sensing data and GIS [12], [13].

Current research is focused on the application of the object based mapping using existing experience, available data and learned methodology in eCognition (Fig.1). The research area includes the eastern part of the Brussels city, Belgium (Fig.2). The data included panchromatic multispectral very high resolution (VHR) image covering study area of Brussels city (Fig.3). Since the construction was performed recently and intensively in the eastern part of the city, the newly constructed built-up areas can be detected here most evidently (Fig.4). For this reason, it has been selected as a research area. Technically, the research was performed in eCognition raster processing software, which enabled to test raster image processing using multiresolution segmentation approach.

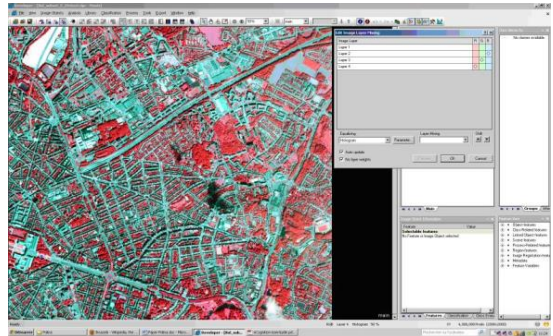

Figure 3. Processing VHR panchromatic image loaded in eCognition

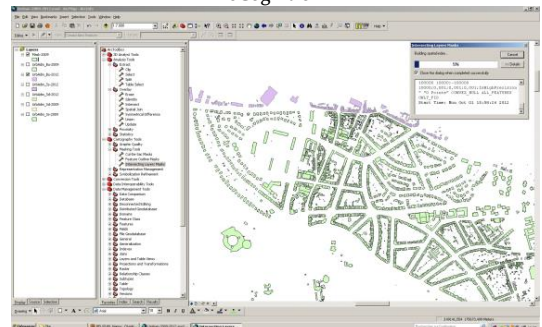

Figure 4. Intersection of vector layers in ArcGIS: recently constructed buildings (lilac) and already existing ones (green) for change detection.

## II. ORGANIZING PROJECT IN ECOGNITION

### A. Importing raster image of the study area

After the image was imported into eCognition software, it was stored in the eCognition project file (.dpr), which was managed in a workspace using guides [1]. The image was then processed using multiresolution segmentation tool in eCognition. During this procedure it was segmented into the defined areas and urban objects were recognized (Fig.5).

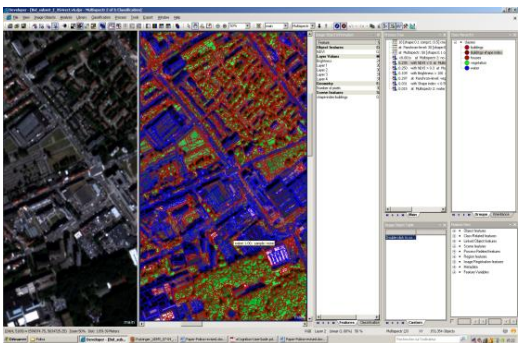

Figure 1. Image classification in eCognition software. Left: initial raw image. Right: classified image

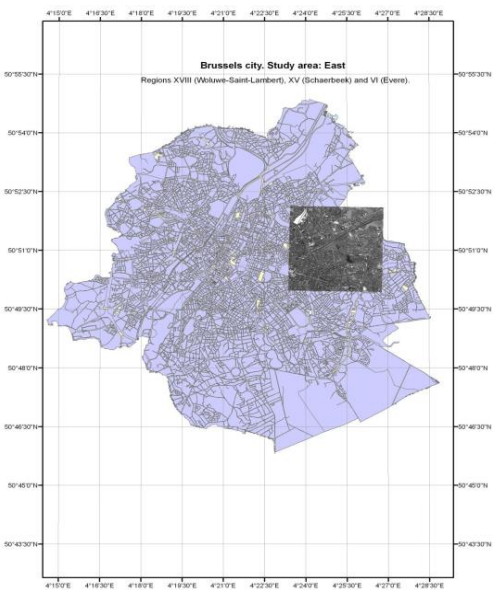

Figure 2. Study area: Brussels, Belgium

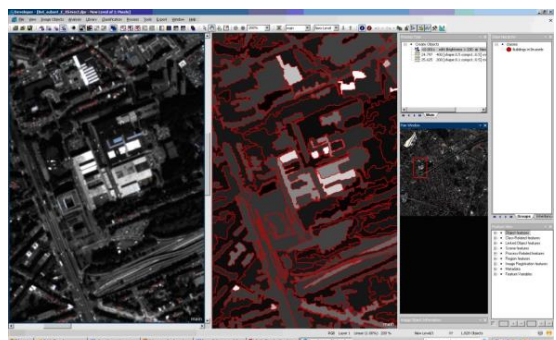

Figure 5. Level of details in classifying panchromatic image, eCognition.

### B. Object oriented mapping

Unlike traditional per-pixel methods of classifications, object-oriented method is based on clustering image into the homogeneous pixels

(objects) and classifying these using their properties: spectral, spatial, textural, relational and contextual ones. Instead of treating image as a combination of pixels that are to be classified on their individual spectral (per pixel) properties, they are grouped into the segments, or clusters.

Thereafter, these segments are to be classified according to their spectral and other criteria: topology (i.e. relationship to the neighboring objects), scale, homogeneity, shape, size. The great advantage of the object based analysis consists in the fact that in such a way, objects with heterogeneous reflectance values, such as building roofs, tree canopy, parts of the road, can be easily recognized despite their heterogeneity. The attribute information to assist this recognition can be taken from the existing geospatial databases. The cues used for the extraction of information play important role in the object oriented analysis since they enable to detect objects that belong to certain classes more effectively. Landscape metrics calculated on the basis of the per-pixel classified images have been extensively used to quantify land use patterns and relate them to the ecological or geographical processes [9], as shape plays important role in spatial attributes of the land use segments in the remotely sensed imagery.

However, it is not often used as a feature in land use classification due to still popular and easy-to-do per-pixel methods of classifications. The example of the integrated methods of fuzzy logic with landscape metrics or per-pixel with object based classifications are demonstrated a systematic way to derive comparable measures of change in urban neighborhoods [16], [2]. For instance, knowledge of the approximate position and class of the roads or buildings decreases processing time, possible errors or misclassified objects. The a priori knowledge can be included in the map as a probability that the land cover change will happen in accordance to the general landscape pattern on the previous maps, e.g. the same territory in different time span (Fig.6).

User guidance enables to define precisely the system components and object elements. Thus, by too detailed scales the machine separated slopes of the same roof as two separate objects due to the slight distinctiveness of the pixels spectral values, while they belong to the same objects (a roof) and separate every single tree while recognizing park and garden areas. Therefore, to avoid such logical unclearness, I applied several trial scale settings of the eCognition scale parameters (30, 50, 70 and 100). These has been illustrated below in order to compare various results that demonstrate impact of scale factor.

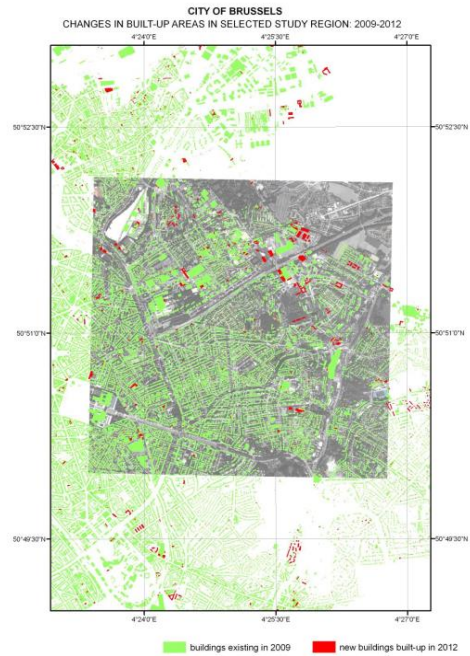

Figure 6. Intensification of the building constructions (ArcGIS visualization)

## II. Multiresolution Segmentation of the Multispectral image

The next step included multiresolution segmentation of the multispectral image. During this procedure, the image has been divided into regions, or segments. Several scale factors have been tested to analyze results. As the scale factor impacts results much, several trial combinations have been tested.

### A. Multiresolution Segmentation for Multispectral image, scale 30.

Initially, the parameters for the segmentation have been set up and at the first time chosen as following ones: scale 1:30, shape homogeneity criterion 0.1. (Fig.7) However, there were some drawbacks arose at scale 30 (Fig.8) and I tested then scale 50.

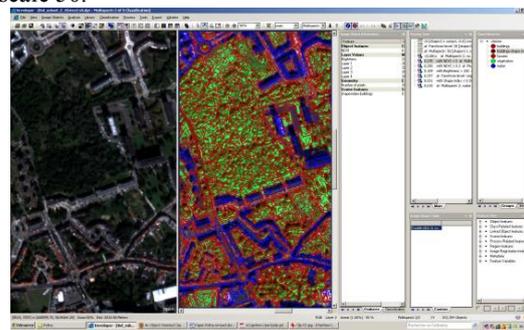

Figure 7. Results of the multiresolution segmentation, scale 30

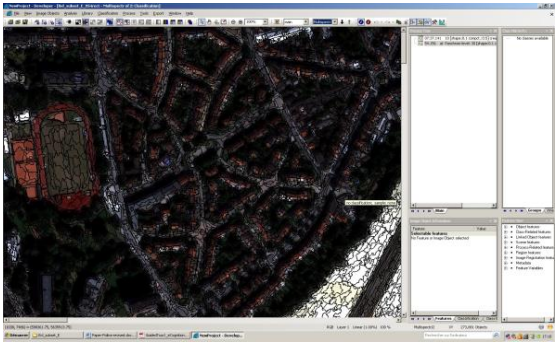

Figure 8. Example of the too detailed scale while classifying park area: every single tree is classified separately

*B. Multiresolution Segmentation for Multispectral image, scale 50.*

The appropriate parameters for the segmentation have been set up several times and chosen as following ones: scale 1:50, shape homogeneity criterion 0.1 (Fig.9).

Various scale influence outcomes of the segmentation significantly.

The appropriate parameters for the segmentation have been set up several times and chosen as following ones: scale 1:50, shape homogeneity criterion 0.1 (Fig.10).

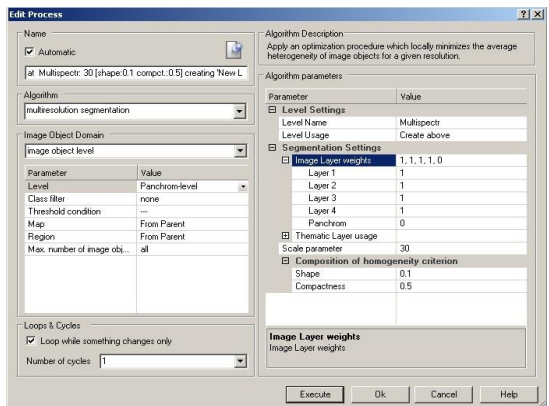

Figure 9. Setting up parameters for Multiresolution Segmentation, scale 30

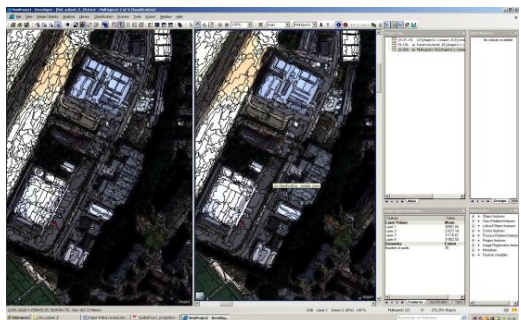

Figure 10. Results of Multiresolution Segmentation of the Multispectral image, scale50.

### III. Detecting Impervious structures within the city area using brightness properties

The final step included detection of buildings within the city area. The recognition has been tried out using brightness properties and shape index. The shape index indicates the general form of the objects (e.g. circular of rectangular). The shape index can be created using formula "Shape index=$1.27AL2$", where A = represents the area of shape in km2, and L = is the length of the longest axis in km. For example, the value of 1.0 expresses circular shape. As long as the shape is elongated, the value of the index became lower. For the buildings the shape was tried out at the area of 0.07 – 0.55. Another parameter for buildings detection is brightness value.

After trying out and testing various scale and homogeneity parameters, scale 50 was defined as the most optimal for research aim: to identify buildings neither too detailed to divide parts of them nor too rough to merge several objects in one. The segmentation with scale 50 gave optimal results, and recognized image objects with sufficient level of minuteness, though more logically accurate than in case of the other scales.

As a result of the image segmentation image objects have been created. These objects have been classified in the next working step. After classes were assigned (vegetation areas, water bodies and buildings), the classification has been performed using standard procedure in eCognition

### *IV. Limitations*

Remote sensing data analysis involves uncertainty caused by limitations of the data and the image interpretation methods used. Since uncertainties accumulate through the processing chain, they will affect final maps inferred from the initial remote sensing data and the derived patterns, quantified by means of spatial metrics and methods [10]. Alike other methods, the OBIA approach may be not suitable to the level of details of the target objects (Fig.11).

Thus, for example, the features or their parts can be not visible on the image. Another example is not corresponding spatial resolution of the map and that of an image (map is more precise than raster image or vice versa). Finally, some additional errors may be caused by the false topological relations of the objects that appear during the generalization. And some problems arising while overlapping and labeling objects should be mentioned as well [11]. Some uncertainness still to mention: shadows from the buildings, directions of roofs, etc.

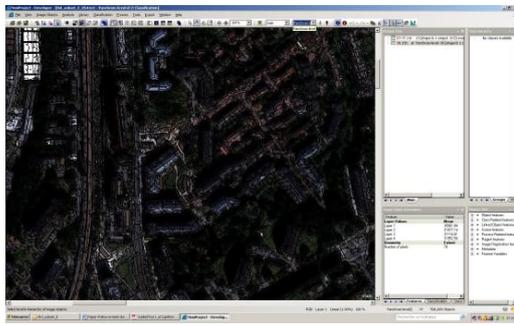

Figure 11. Results of Multiresolution Segmentation of the panchromatic image, scale 30, fragment.

### V. Conclusion

This paper contributed towards the experience of GIS based high precision urban mapping. Particularly, the research detailed application of the multiresolution segmentation towards the image analysis for urban studies with a case study of Brussels. It demonstrated highly effective and useful application of the eCognition software for mapping and visualization urban land cover types. It furthermore proved that urban mapping using very high resolution (VHR) satellite images is indeed effective using object based image analysis approach (Fig.12).

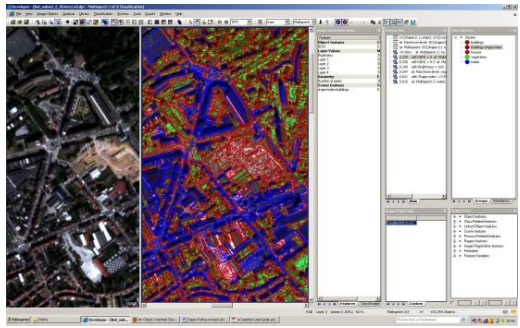

Figure 12. Results of segmentation: recognized objects. Linear structures – roads, streets (blue), park spaces (green), contours of buildings (red).

To conclude, the OBIA and eCognition software can be successfully applied towards effective cadastral mapping and is specifically recommended for urban studies.